\documentclass[reprint,amsmath,amssymb,prl,aps,showkeys,showpacs]{revtex4-1}
\usepackage{graphicx}
\usepackage{dcolumn}
\usepackage{bm}
\usepackage{float}

\begin{document}
\title{A sidewall friction driven ordering transition in granular channel flows: Implications for granular rheology}
\author{Sandip Mandal}
\author{D. V. Khakhar}\email{khakhar@iitb.ac.in}
\affiliation{Department of Chemical Engineering, Indian Institute of Technology Bombay, Powai, Mumbai 400076, India}

\date{\today}

\begin{abstract}
We report a transition from a disordered state to an ordered state in the flow of nearly monodisperse granular matter flowing in an inclined channel with a bumpy base, in discrete element method simulations. For low particle-sidewall friction coefficients, the particles are disordered and the \textit{Bagnold} velocity profile is obtained. However, for high sidewall friction, an ordered state is obtained, characterized by a layering of the particles and hexagonal packing of the particles in each layer. The extent of ordering, quantified by the local bond-orientational order parameter, varies in the cross-section of the channel, with the highest ordering near the side walls. The flow transition significantly affects the local rheology -- the effective friction coefficient is lower, and the packing fraction is higher, in the ordered state compared to the disordered state. A simple model, incorporating the extent of local ordering, is shown to describe the rheology of the system.   
\end{abstract}


\maketitle
Understanding the flow of granular materials is important in the context of natural phenomena and industrial processes \citep{van1984sediment, andreotti2002selectiona, hutter1995dynamics, perry1997,cleary2016}. Considerable progress has been made in the development of theories for the rheology of granular flows, with the objective of developing continuum models for the analysis and design of large systems \citep{campbell90, goldhirsch03, nedderman2005statics, forterre08, rao2008}. Granular materials exhibit many complex phenomena \cite{jaeger96,ottino00}. The ordering of monodisperse granular materials, when subjected to vibration or shear flow, is one example \citep{pouliquen1997crystallization, olafsen1998clustering, tsai2003internal, tsai2004slowly, daniels2005hysteresis, reis2006crystallization}. The transition from a disordered state to an ordered state is marked by a significant increase in the solid fraction manifested as a compaction of the material \citep{pouliquen1997crystallization,daniels2005hysteresis}. In sheared systems, which are the focus of the present study, ordered states show a layering of particles with adjacent particle layers sliding past each other without much interchange of particles between the layers \citep{ silbert2002boundary, tsai2003internal, tsai2004slowly}. Shear stresses are consequently lower compared to the disordered state. Order-disorder transitions thus affect system behavior significantly, however, they have not been explicitly incorporated in the analysis of granular rheology.

\begin{figure}
\includegraphics[width=1.8in]{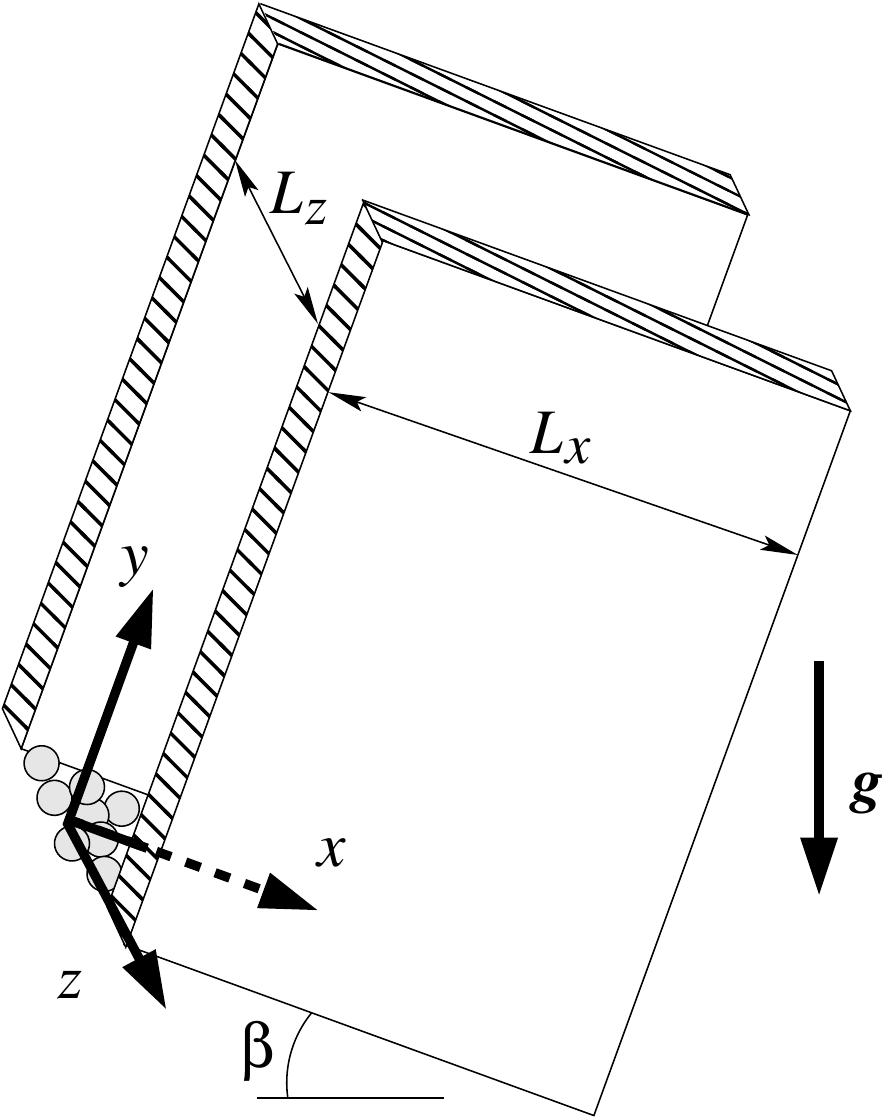}
\caption{\label{fig:schem} A schematic view of the system geometry showing the coordinate system and the direction of gravity  ($\boldsymbol{g}$). The channel (length $L_x$ and width $L_z$) has frictional sidewalls, a bumpy base and is inclined at an angle $\beta$. The flow is in the $x$ direction.} 
\end{figure}

The most detailed studies of order-disorder transitions in granular shear flow are computational, using the discrete element method (DEM) to simulate the flow of monodisperse particles (diameter $d$) on a rough inclined plane with periodic boundary conditions (no sidewalls)  \citep{silbert2002boundary, kumaran2012transition, kumaran2013effect}. \citet{kumaran2012transition} carried out a study of the transition using a base comprising particles of diameter, $d_b$, arranged on a square lattice. When the base roughness parameter ($d_b/d$) was less than 0.6, the system underwent a transition to an ordered state, comprising hexagonally close packed particle layers sliding on each other. The shear stress ($\tau_{xy}$) in the ordered state followed the Bagnold scaling ($\tau_{xy}\propto\dot{\gamma}^2$, where $\dot{\gamma}$ is the shear rate) but with smaller values of the Bagnold coefficients compared to the disordered state. The transition was very sharp, occurring over a 1\% change in the roughness parameter. A similar ordering was seen in experimental studies of glass beads flowing in an annular shear cell with smooth sidewalls by \citet{tsai2003internal, tsai2004slowly} and \citet{savage93}. \citet{hill2003structure} and \citet{bi2006experimental} reported layering in the flowing layer in a quasi-2d rotating cylinder and an inclined chute, respectively, but did not examine the ordering within the layers. \citet{orpe2009fast} showed local ordering near the sidewalls for the flow in a vertical channel. A large number of studies of dense granular flows in narrow channels have been carried out in the  past\citep{rajchenbach1990, khakhar01a, komatsu2001creep, orpe2001scaling, jain2002, taberlet2003superstable, hill2003structure, orpe2004solid,  jop2006constitutive, bi2006experimental, orpe2007rheology, fan2013}. While the effects of sidewalls on the flow have been previously noted \citep{jop2005crucial}, their role in initiating ordering has not been examined. In this work, we study the gravity driven flow of nearly monodisperse particles in a rectangular channel with a rough base and planar frictional sidewalls (Fig.~\ref{fig:schem}), by means of DEM simulations,  to understand the factors affecting the transition to ordered states and to characterize the rheology of the flow. 

The system comprises an inclined channel having planar sidewalls with a particle-wall friction coefficient, $\mu_{pw}$, and a bumpy base made up of a 1.2$d$ thick slice of a random close packed bed of mono-sized spheres of diameter $d$ (Fig.~\ref{fig:schem}). The channel has a length $L_x=20d$ and width $L_z=10d$, with periodic boundary conditions on the surfaces normal to the $x$-axis. $N\in(3000,10500)$ particles of diameter $d$ and mass $m$ are used in the simulations, with a polydispersity of $\pm5$\% in the particle diameter. The contact forces generated due to the deformation of the spheres are calculated using the L3 model of \citet{silbert2001granular}. The equations of motion are non-dimensionalized using $d$, $m$, $mg$, and $\sqrt{d/g}$ as length, mass, force, and time units and all results are presented in terms of dimensionless variables. The detailed simulation methodology is given in \cite{tripathi2011rheology}. The parameters used in the simulations are: particle-particle friction coefficient, $\mu_p=0.5$, coefficient of restitution, $e=0.88$, dimensionless normal spring constant, $k_n=2\times10^5$, and tangential to normal spring constant ratio, $k_t/k_n=2/7$.

The flow achieves a steady state only for a range of inclination angles ($\beta$) and all results reported here are averaged over 4 sets, each with a time duration of 1000 units at the steady state. For coarse graining, we use bins with dimensions $20\times 1\times1$ to calculate the mean velocity ($v_x$), volume fraction ($\phi$, the volume of particles in a bin divided by the volume of the bin), and total stress tensor ($\bm{\sigma}$) composed of streaming and collisional components computed as in Ref.~\cite{mandal2016study}. The deviatoric stress tensor ($\bm{\tau}$) is given by $\bm{\tau}=-\bm{\sigma}-P\bm{\delta}$, where $\bm{\delta}$ is the unit tensor and $P=-tr(\bm{\sigma})/3$ is the pressure. The shear rate ($\dot{\gamma}=dv_x/dy$) is obtained by differentiating the mean velocity profile using the forward difference method. The number density ($n$) is calculated based on the positions of the centroids of the particles using bins with a small height (dimensions $20 \times 0.05\times1$) so as to probe layering in the system. The local bond-orientational order parameter ($q_6$) in the $x$--$z$ plane is calculated from 
\begin{equation}
q_6=\Big|\frac{1}{N_b}\sum_{j=1}^{}{\frac{1}{N_{bj}}\sum_{k=1}^{N_{bj}} {\exp(i6\theta_{jk})}}\Big|, 
\end{equation} 
where $N_b$ is the number of particles in a bin, $N_{bj}$ are the  nearest neighbors of particle $j$, and $\theta_{jk}$ is the angle made with the $x$ axis by the line joining the centroids of the reference particle $j$ and its neighbor $k$. Since the particles are layered in the ordered state, nearest neighbors of the reference particle are taken to be only those particles which are at radii smaller than the first minimum in the radial distribution function and at heights $\pm 0.1d$ from the height of the reference particle. $q_6=1$ for a hexagonally ordered system and $q_6=0$ for a completely disordered system. 

\begin{figure}
\includegraphics[width=2.5in]{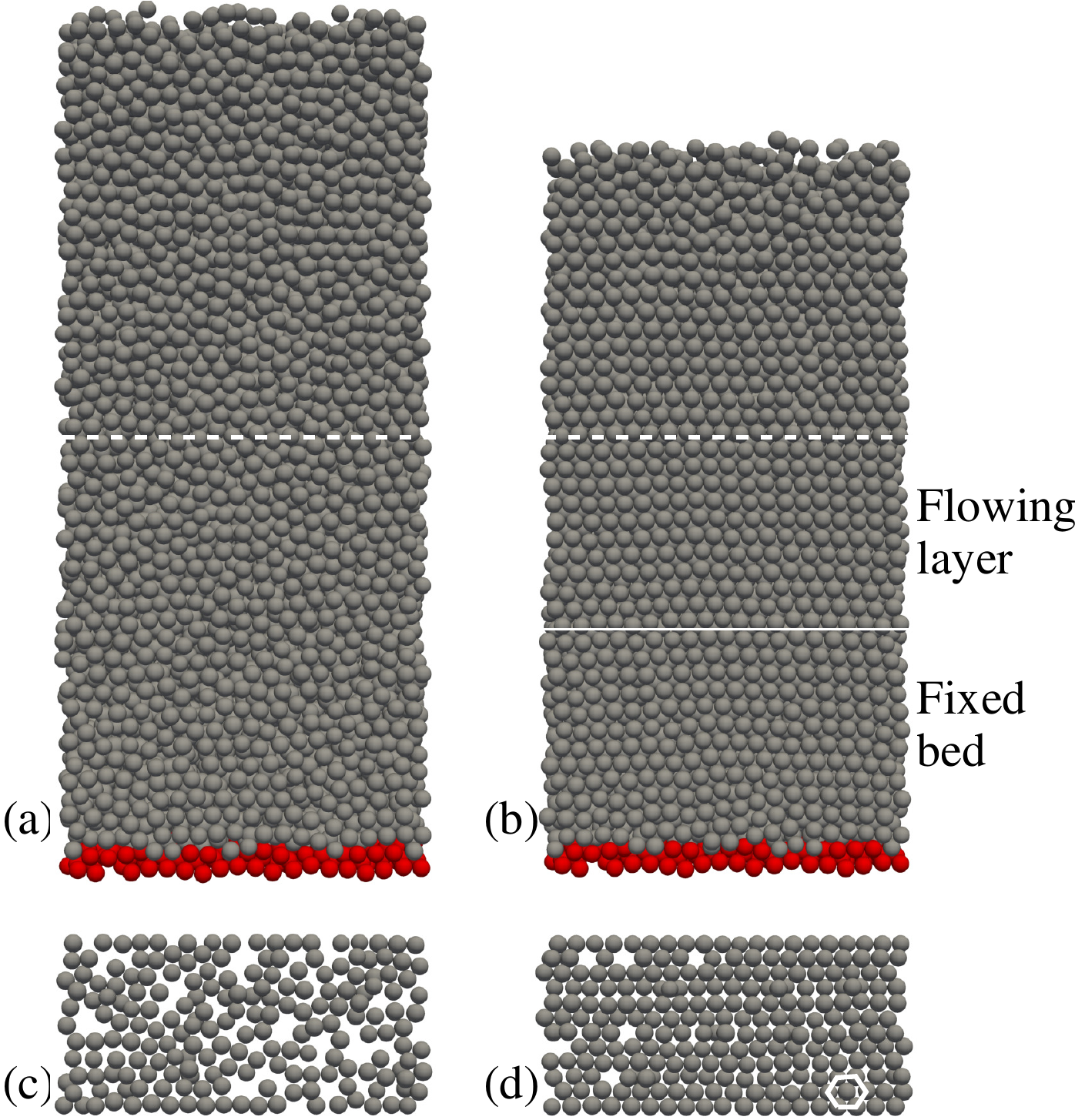}
\caption{\label{fig:order-disorder} Snapshots of the system for $N=9000$ flowing particles for the channel inclined at $\beta=28^\circ$. Side view of the system for (a) low particle-wall friction coefficient ($\mu_{pw}=0.0001$) and (b) high particle-wall friction coefficient ($\mu_{pw}=0.15$). A fixed bed is formed for $\mu_{pw}=0.15$ and the interface between the fixed bed and the flowing layer is shown by a solid line in (b). Particles in a slice of thickness $1d$ taken at the dashed lines for (a) and (b) are given in (c) and (d), respectively. Line shows the hexagonal ordering in (d).} 
\end{figure}

Fig.~\ref{fig:order-disorder}, which shows snapshots of the system at steady state, qualitatively illustrates the effect of wall friction on ordering. When the sidewalls are nearly frictionless ($\mu_{pw}=0.0001$, Fig.~\ref{fig:order-disorder}a), the particles are disordered, whereas, for highly frictional walls ($\mu_{pw}=0.15$, Fig.~\ref{fig:order-disorder}b), the particles near the base form a fixed bed and the particles are arranged in layers. The system also shows significant compaction and the height decreases by $\sim7d$ for the latter case. A slice of thickness one particle diameter taken in the flowing region (dashed line in Fig.~\ref{fig:order-disorder}b) shows hexagonal ordering (Fig.~\ref{fig:order-disorder}d) in the case of high $\mu_{pw}$, however, no in-plane ordering is observed in the case of low $\mu_{pw}$ (Fig.~\ref{fig:order-disorder}c). A video showing the flow in the two states is included in the supplementary material of the paper.

\begin{figure}
\includegraphics[width=8.6cm]{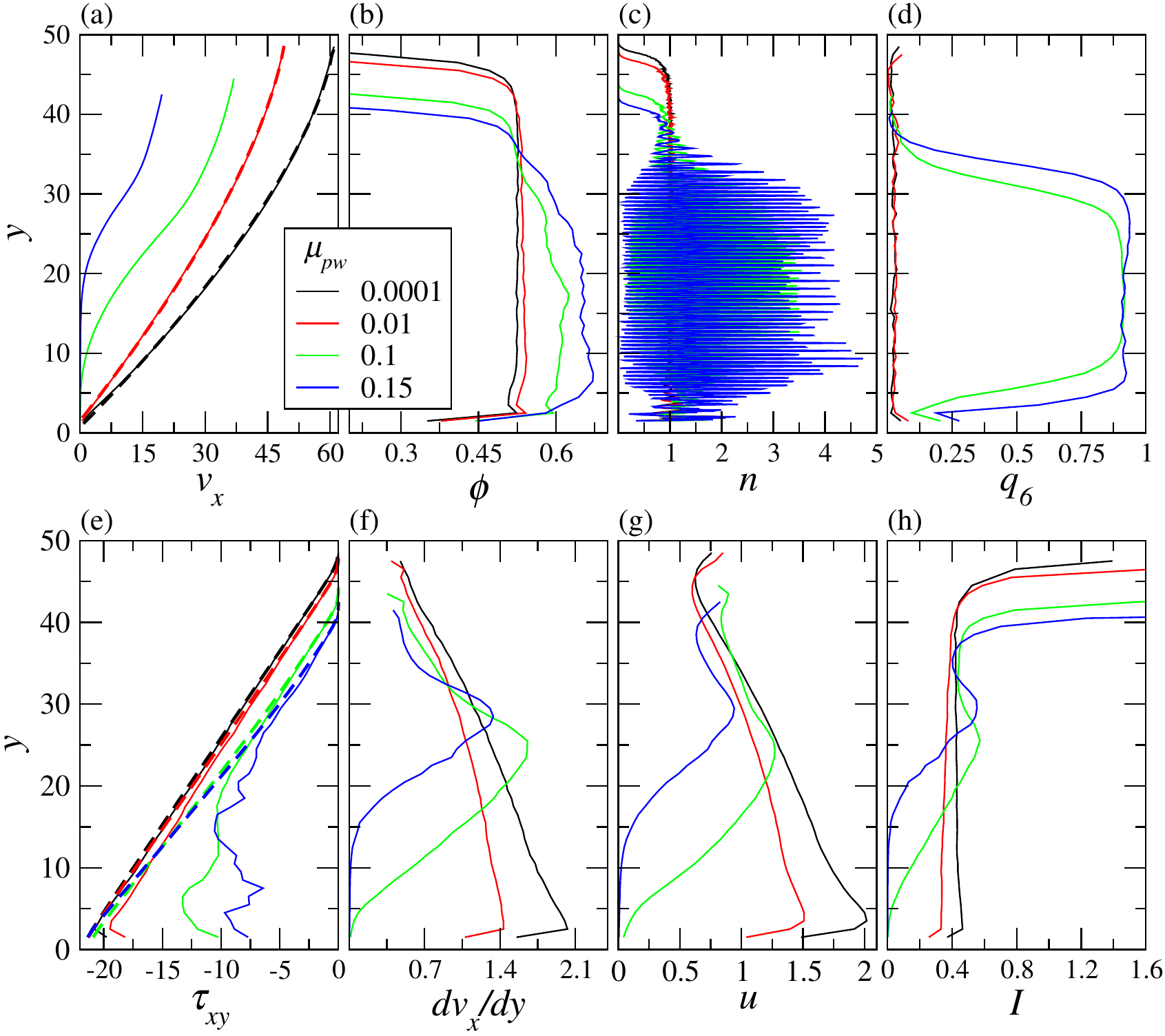}
\caption{\label{fig:vx-muw} (a) Mean velocity ($v_x$), (b) volume fraction ($\phi$), (c) number density ($n$), (d) local bond-orientational order parameter ($q_6$), (e) shear stress ($\tau_{xy}$), (f) shear rate ($dv_x$/$dy$), (g) root mean square velocity ($u$),  and (h) inertial number ($I$) profiles obtained from the flow of 9000 spheres down the channel inclined at $\beta=28^\circ$ for various particle-wall friction coefficients ($\mu_{pw}$) indicated in the legend. The dashed lines in (a) are \textit{Bagnold} velocity profiles and in (e) are predictions of the momentum balance equation.} 
\end{figure} 

Fig.~\ref{fig:vx-muw} shows the effect of the particle-wall friction coefficient ($\mu_{pw}$) on mean velocity ($v_x$), volume fraction ($\phi$), number density ($n$), local bond-orientational order parameter ($q_6$), shear stress ($\tau_{xy}$), shear rate ($dv_x/dy$), root mean square velocity ($u$), and inertial number ($I=\dot{\gamma}d/(P/\rho_p)^{1/2}$, where $\rho_p$ is the particle density) profiles, at the center-line of the channel ($z=0$). With increasing $\mu_{pw}$, the maximum velocity reduces and a fixed bed forms near the base. For the low friction cases ($\mu_{pw}=0.0001,0.01$) the velocity profiles closely match the \textit{Bagnold} velocity profile (dashed lines, Fig.~\ref{fig:vx-muw}a). The volume fraction profile along the layer depth is uniform for low $\mu_{pw}$ but increases with the depth at higher $\mu_{pw}$. The magnitude of the volume fraction increases significantly with increasing $\mu_{pw}$ as discussed above. The number density profile (Fig.~\ref{fig:vx-muw}c), obtained using bins of height 0.05, clearly shows a layering of the particles (indicated by high local values of $n$) at the higher values of $\mu_{pw}$ and no layering at lower values. A near-perfect in-plane hexagonal ordering ($q_6 \approx 1$) is observed in the middle region of the flowing layer at higher $\mu_{pw}$ (Fig.~\ref{fig:vx-muw}d), however, the material is in a disordered state near the free surface. The extent of ordering increases and the ordering front propagates to greater heights with increasing $\mu_{pw}$ (Fig.~\ref{fig:vx-muw}d). The shear stress increases linearly with increasing depth in the flowing region and matches with the momentum balance prediction for the case of frictionless sidewalls (dashed lines, Fig.~\ref{fig:vx-muw}e). However, its variation is complex and magnitude is lower in the fixed bed regions for the cases at higher $\mu_{pw}$. For low $\mu_{pw}$, the shear rate increases monotonically with increasing depth; whereas, for higher values of $\mu_{pw}$, the shear rate exhibits a maximum and goes to zero in the fixed bed, as seen previously in experimental results for channel flows \cite{orpe2004solid}. The root mean square (\textit{rms}) velocity ($u$) exhibits a variation qualitatively similar to the shear rate. The inertial number is nearly constant over the depth of the layer for the low wall friction cases, but shows a maximum at higher $\mu_{pw}$. Increasing the flowing layer height ($H$) by increasing the number of particles ($N$), keeping other parameters fixed, results in a similar disorder-order transition. Reduction in inclination angle ($\beta$) also causes the transition. The disorder-order flow transition is robust and is observed for a wider channel ($L_z=20$) and when the poly-dispersity is increased to $\pm10$\% of the diameter, though the extent of ordering is reduced in the latter case.

\begin{figure}
\includegraphics[width=1.5in]{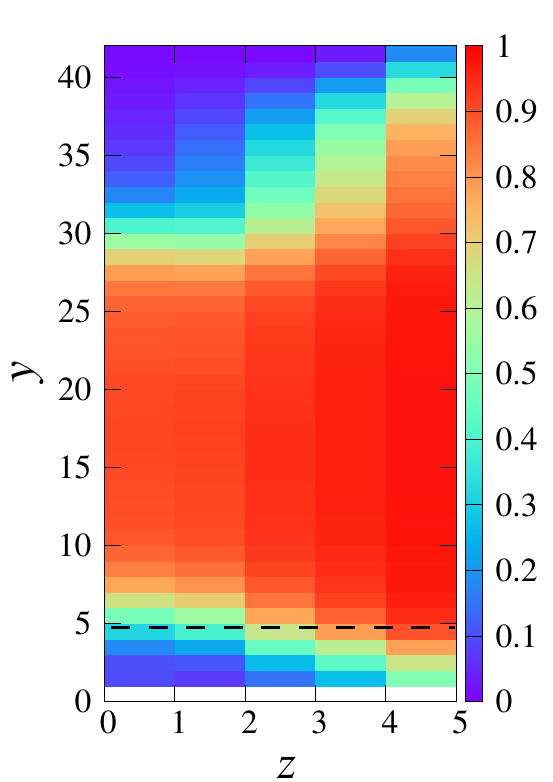}
\caption{\label{fig:q6}Variation of $q_6$ in the $y$--$z$ plane in the case of $\mu_{pw}=0.10$, $N=9000$, and $\beta=28^\circ$. The dashed line shows the interface between the fixed bed and the flowing layer.} 
\end{figure}

Fig.~\ref{fig:q6} shows the variation of the order parameter, $q_6$, in the cross-section of the channel for one case. The results indicate that the extent of ordering varies in the cross-section, with the highest degree of ordering near the sidewalls ($z=5$)  and the lowest  near the base ($y=0$) and the free surface ($y=H$). The results presented above indicate the following mechanism: when the wall friction becomes large enough, a partially ordered fixed bed is formed, which presents a smoother base for the flow and promotes layering. Ordered regions are then nucleated at the sidewalls and grow inwards with increasing wall friction. The transition in the present case is also boundary driven, as found previously for a system without sidewalls \cite{kumaran2012transition}.

\begin{figure}
\includegraphics[width=3.4in]{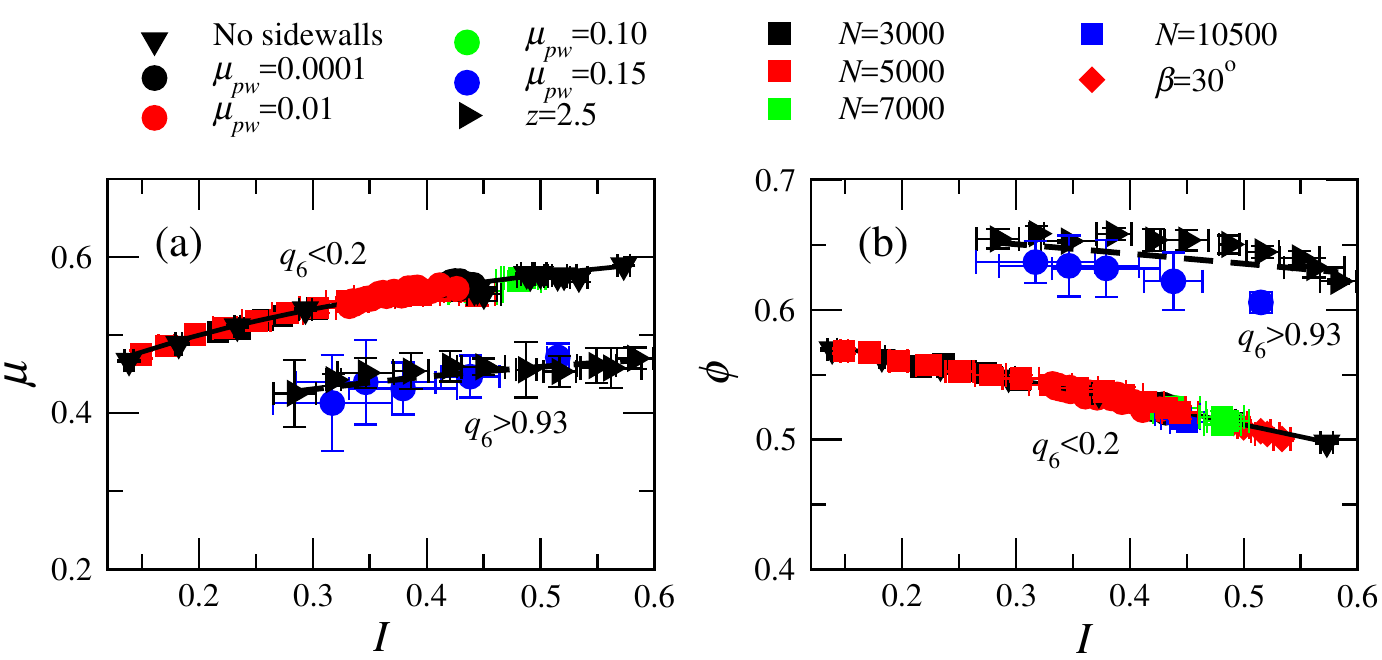}
\caption{\label{fig:mu-I}Variation of (a) effective friction coefficient ($\mu$) and (b) volume fraction ($\phi$) with inertial number ($I$) for different cases. The variation from the base case values  ($\beta=28^{\circ}$, $z=0$, $\mu_{pw}=0.1$, $N=9000$) for each data set is indicated in the legend. Only data in the ordered regions ($q_6>0.93$) and in the disordered regions ($q_6<0.2$) are included. Error bars are standard deviations over four sets. The solid lines and dashed lines are fits of (a) Eq.~\ref{eq:mu} and (b) Eq.~\ref{eq:phi} to the data in the disordered and ordered regions, respectively. The fitted model parameters are listed in Table~\ref{tab:parameters}.} 
\end{figure}

\begin{figure}
\includegraphics[width=8.6cm]{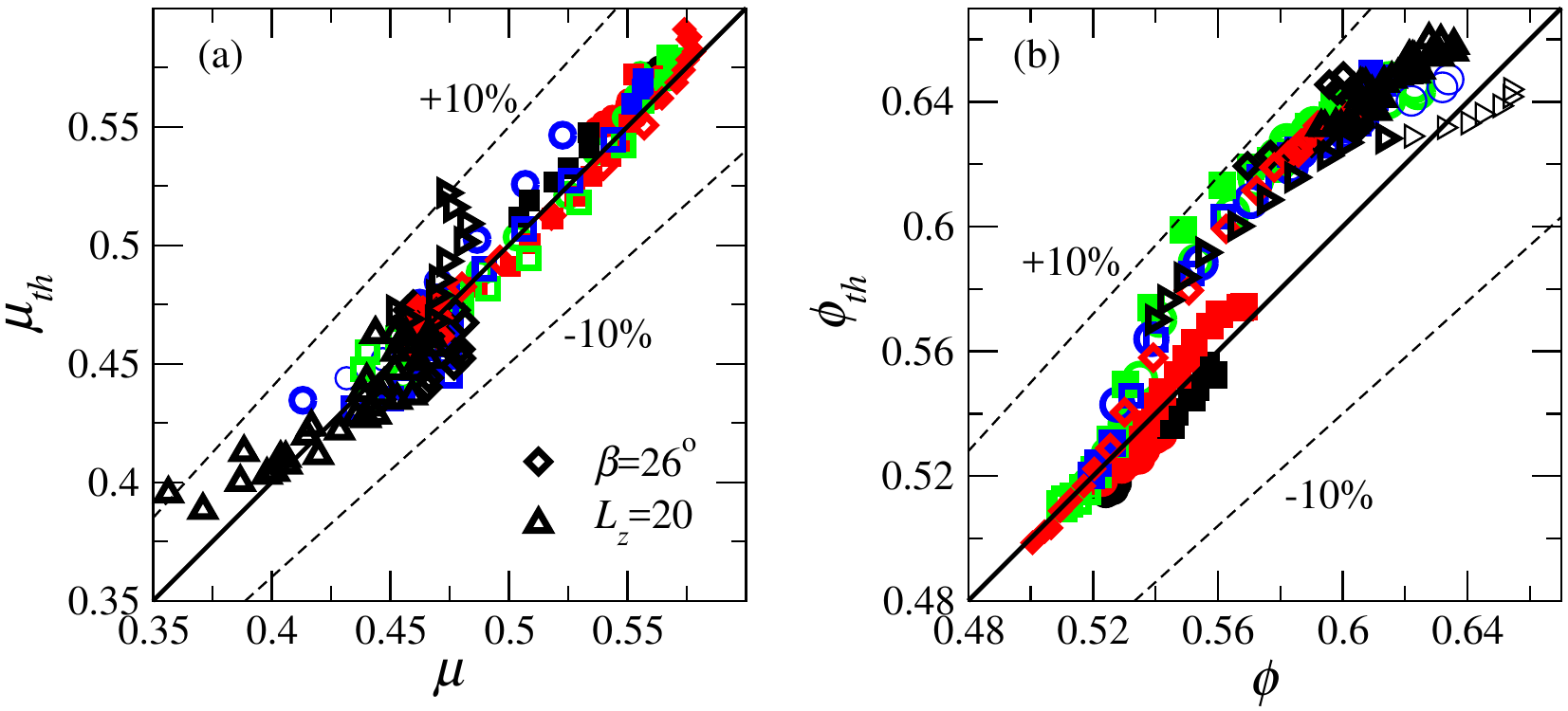}
\caption{\label{fig:model-simulation} Comparison of the estimated (a) friction coefficient ($\mu_{th}$) and (b) solid fraction ($\phi_{th}$) to the simulation results ($\mu,\phi$) for the different cases shown in Fig.~\ref{fig:mu-I}. Data for the partially ordered regions are included along with two additional cases shown in the legend. Open symbols: data in the ordered region ($q_6>0.93$), filled symbols: data in the disordered region ($q_6<0.2$), and thick open symbols: data in the partially ordered region ($0.93>q_6>0.2$).} 
\end{figure}

We examine the rheology of the system in the framework of $\mu$--$I$ and $\phi$--$I$ scaling relations \cite{jop2006constitutive} in Fig.~\ref{fig:mu-I}, where $\mu=|\bm{\tau}|/P$ is the shear rate dependent friction coefficient. A large number of sets in which only one of the parameters is varied from the base case values ($\beta=28^{\circ}$, $z=0$, $\mu_{pw}=0.1$, $N=9000$) are shown for data in either the highly ordered ($q_6>0.93$) or highly disordered ($q_6<0.2$) regions. The data collapse is very good for $\mu$ and the results show a significant reduction in the friction coefficient with ordering. The collapse is also very good for $\phi$ in the disordered regions and ordering results in an increase in volume fraction, but the collapse is not as good for the data in the ordered regions. Data corresponding to intermediate ordering fall between the data clusters shown. Following \citet{jop2006constitutive}, we fit the following empirical equations to the data
\begin{equation}\label{eq:mu}
\mu_{th}=\mu_s+(\mu_m-\mu_s)/(1+I_0/I), 
\end{equation}
\begin{equation}\label{eq:phi}
\phi_{th}=\phi_{max}-aI^b, 
\end{equation}
and fitted values of the model parameters ($\mu_s$, $\mu_m$, $I_0$, $\phi_{max}$, $a$, $b$) are given in Table~\ref{tab:parameters} for the ordered and disordered regions. The fits are quite good for both $\mu$ and $\phi$. To describe the rheology of regions with partial ordering, we assume the model parameters to be linear functions of $q_6$ interpolating between the values for the highly ordered and disordered states (Table~\ref{tab:parameters}). Fig.~\ref{fig:model-simulation} shows a comparison of the predicted values ($\mu_{th},\phi_{th}$) to the simulation results for all the data, including those from partially ordered regions. The predictions are quite good for both $\mu$ and $\phi$ ($\pm10$\%), indicating that the simple model proposed here can describe the local rheology, provided the extent of local ordering is known.

\begin{table}
\caption{\label{tab:parameters} Model parameters of the constitutive equations of $\mu-I$ rheology.}
\begin{ruledtabular}
\begin{tabular}{ccccccc}
$\bar{q}_6$ & $\mu_s$ & $\mu_m$ & $I_0$ & $\phi_{max}$ & $a$ & $b$ \\
\colrule
0.08 & 0.38 & 0.73 & 0.38 & 0.59 & 0.18 & 1.18\\
0.94 & 0.33 & 0.56 & 0.38 & 0.67 & 0.08 & 1.18\\ 
\end{tabular}
\end{ruledtabular}
\end{table}

We show that sidewall  friction results in complex ordering of granular material in channel flows, which causes local compaction and reduction in the local friction coefficient. The ordering is facilitated by the formation of a partially ordered fixed bed, which results in a reduction in the effective roughness of the base. The mechanism is robust and ordering is obtained for wide range of system parameters, including increased polydispersity. The model presented is a first step in the analysis of such partially ordered flows, however, prediction of the local extent of ordering remains a challenge. The results have implications for the analysis of channel flow experiments as well as for modelling such systems.           

\begin{acknowledgements}
The authors acknowledge the financial support of the Science and Engineering Research Board, India through grant SR/S2/JCB-34/2010.
\end{acknowledgements}
\end{document}